\documentstyle[12pt]{article}
\newcommand{\gsim}{\; \mbox{\raisebox{-.05in}{$\stackrel{\textstyle
>}{\sim}$}} \;}
\newcommand{\lsim}{\; \mbox{\raisebox{-.05in}{$\stackrel{\textstyle
<}{\sim}$}} \;}
\begin{document}
\begin{center}
{\bf Superconductivity-Induced Anomalies in the Spin Excitation 
Spectra of Underdoped $\bf YBa_2 Cu_3 O_{6+x}$}

\vspace{0.5in}

H.F. Fong and B. Keimer 

\vspace{0.1in}

Department of Physics, Princeton University, Princeton, NJ 08544\\
and Brookhaven National Laboratory, Upton, NY 11973

\vspace{0.2in}

D.L. Milius and I.A. Aksay

\vspace{0.1in}

Department of Chemical Engineering, Princeton University, Princeton, NJ
08544\\
\end{center}

\vspace{1.2in}

\begin{center}
ABSTRACT
\end{center}

\vspace{.15in}

\noindent Polarized and unpolarized neutron scattering has been used to
determine the effect of superconductivity on the magnetic excitation
spectra of $\rm YBa_2 Cu_3 O_{6.5}$ ($\rm T_c = 52$K) and $\rm 
YBa_2 Cu_3 O_{6.7}$ ($\rm T_c = 67$K). Pronounced enhancements of the
spectral weight centered around 25 meV and 33 meV, respectively, are
observed below $\rm T_c$ in both crystals, compensated predominantly by a loss
of spectral weight at {\it higher} energies. The data provide important
clues to the origin of the 40 meV magnetic resonance peak in
$\rm YBa_2 Cu_3 O_7$.

\vspace{.25in}


\vspace{.2in}

\clearpage

\pagestyle{plain}

Recently a new magnetic resonance mode has been found in the
superconducting state of $\rm YBa_2 Cu_3 O_7$ by inelastic neutron
scattering. The mode occurs at an energy of 40 meV 
and wavevector ${\bf q} = (\frac{\pi}{a},\frac{\pi}{a})$, and disappears
in the normal state \cite{fong,bourges,mook}. Since this mode is a novel
signature of the superconducting state in the cuprates and has
not been observed in any other material, much theoretical
effort has gone into its interpretation.
Qualitative considerations on the basis of a BCS
pairing model show that quasiparticle pair production across the
superconducting energy gap can give rise to enhanced spin-flip neutron
scattering below the superconducting transition, provided that the order
parameter changes phase on the Fermi surface \cite{fong,scalapino}. 
More elaborate theories are
required to explain the experimental features of the mode in detail,
especially its sharpness in both energy and momentum. Final-state
interactions between the quasiparticles 
\cite{levin,onufrieva,zha,millis-monien}
as well as bandstructure anomalies \cite{bulut,klein}
have been suggested to account for the experimental data in the
framework of a $d$-wave BCS model. 
Predictions derived from the interlayer pair tunneling theory of
high temperature superconductivity explain most features of the neutron data 
without invoking such effects \cite{chakravarty}.
A more complicated order parameter with a sign change between different 
bands \cite{yakovenko} and a collective mode in the
particle-particle (rather than particle-hole) channel \cite{zhang}
have also been proposed. Clearly, more experimental information is required 
in order to discriminate between these divergent interpretations of the 
40 meV mode. Open questions questions include the
dependence of the energy, width, and spectral weight of the
resonance peak on the doping level, and the relation between the
resonance peak and the magnetic excitation spectrum in the normal
state. In fully oxygenated $\rm YBa_2 Cu_3 O_{6+x}$ ($\rm x \sim
1$) the magnetic susceptibility is not strong enough to be
observable by neutron scattering \cite{fong}.

These issues are addressed in this Letter by systematically
studying the influence of superconductivity on the magnetic
excitation spectra of underdoped $\rm YBa_2 Cu_3 O_7$. 
The doping dependence of the magnetic excitation
spectrum of this system has already been characterized extensively 
in prior work. Briefly, the spin excitations in the 
antiferromagnetic regime (x$\lsim$0.4) are well described by spin 
wave theory \cite{tranquada1,reznik,grenoble}. 
For x$\gsim$0.4, the magnetic response is broadened in 
{\bf q} and the spectral weight is depressed at 
low energies \cite{grenoble,tranquada2}. 
We have observed similar features in our samples. However, in these 
second-generation experiments we have taken data with very high counting 
statistics and found that the onset of superconductivity leads to a 
pronounced redistribution of the spectral weight. 
The energy of maximum spectral weight enhancement increases monotonically with
the superconducting transition temperature. 

The samples used for the present study were two $\rm YBa_2 Cu_3 O_{6+x}$
single crystals of volumes $\rm \sim 3$ $\rm cm^3$ and mosaic 
spreads 0.8$^{\rm o}$ and 1.4$^{\rm o}$
(full width at half maximum), respectively. The samples were oxygen-depleted
in their as-grown state and were annealed in air at 
640$^{\rm o}$C for different lenghts of time. The average oxygen content 
during the anneal was monitored by thermogravimetric analysis. After the 
average oxygen content was adjusted by this procedure, the samples were 
sealed in quartz tubes and kept at 740$^{\rm o}$C for about two weeks in order
to achieve a homogeneous distribution of the oxygen content. The uniform
susceptibility measured on a piece cut from the first sample, shown 
in the inset of Fig. 2, demonstrates that the oxygen content is indeed 
highly homogeneous: The width of the superconducting transition 
($\rm T_c = 52K$ midpoint) is very similar to  
the ones reported for very high quality small crystals in this doping regime
\cite{altendorf}. Susceptibility measurements on the second crystal 
revealed $\rm T_c =67$K (midpoint) with a somewhat larger width. 
A comparison with previously reported calibrations of the
lattice constants and transition temperatures \cite{altendorf,cava}
shows that the oxygenation states of the two crystals
are $\rm x \sim 0.5 \pm 0.05$ and $\rm x \sim 0.7 \pm 0.05$, respectively.

The neutron scattering measurements were performed at the H4M, H7 and H8
triple axis spectrometers at the High Flux Beam Reactor at the Brookhaven
National Laboratory. The beam collimations were 40'-40'-80'-80' and the
final energy was 30.5 meV for the unpolarized-beam measurements. The (002)
reflection of pyrolytic graphite (PG) was used as both monochromator and
analyser, and a PG filter was placed before the analyser in order to
eliminate higher-order contamination of the incident beam. For the
polarized-beam measurements we used Heusler alloy crystals as monochromator
and analyser, beam collimations 40'-60'-80'-80' and
28 meV final energy. The flipping ratio was 33 (corresponding to $\sim$95\%
beam polarization) for both horizontal and vertical guide fields at the 
sample. The energy resolutions were $\sim$7 meV 
(full width at half maximum) in both cases. 

Fig. 1 shows typical constant-energy scans for $\rm YBa_2 Cu_3 O_{6.5}$ 
taken with both unpolarized and polarized beams. As previously reported
\cite{grenoble,tranquada2}, the intensity is peaked at an in-plane
wavevector of ${\bf q} = (\frac{1}{2},\frac{1}{2})$. [The reciprocal space
coordinates (H,K,L) are quoted in units of the reciprocal lattice vectors 
$2\pi/a \sim 2\pi/b\sim 1.63 {\rm \AA}^{-1}$ and $2\pi/c\sim 0.53
{\rm \AA}^{-1}$]. At all temperatures the magnetic
intensity is sinusoidally modulated as a function of the wavevector
perpendicular to the $\rm CuO_2$ planes, a behavior 
which is also characteristic of both
acoustic spin waves in antiferromagnetic $\rm YBa_2 Cu_3 O_{6+x}$
\cite{tranquada1,reznik,grenoble} and of the 40 meV 
resonance peak in $\rm YBa_2 Cu_3 O_7$
\cite{fong,bourges,mook}. Our data were taken at momentum transfers for
which this intensity modulation is maximum.

Constant-energy scans with similar counting statistics were taken at a series
of temperatures and fitted to Gaussians in order to extract the amplitude and
width of the magnetic signal. The fitted {\bf q}-width does not change with
temperature outside of the experimental error. By contrast, the amplitude
displays a dramatic temperature dependence (Fig. 2): As previously observed
\cite{grenoble,tranquada2}, the intensity increases gradually as $\rm T_c$
is approached from above. The ensuing abrupt increase by $\sim 30$\% upon
cooling through $\rm T_c$ is only clearly discernible due to the high
counting statistics of our data and has thus far not been reported by other
groups. (A sharp magnetic feature at $\hbar \omega = 27$ meV was previously
observed by Tranquada {\it et al.} \cite{tranquada2} at this doping level but
its temperature dependence was not measured.)
It provides the first direct link between superconductivity
and magnetic excitations of energies comparable to the superconducting 
energy gap in underdoped $\rm YBa_2 Cu_3 O_{6+x}$.

In principle, the observed enhancement of the neutron cross section below 
$\rm T_c$ could also arise from a superconductivity-induced phonon shift and/or
linewidth change. Such effects are known to be negligible for ${\bf q}=0$
phonons in underdoped $\rm YBa_2 Cu_3 O_{6+x}$ 
\cite{altendorf,cardona}, but since the observed
effect is rather subtle we used two independent methods to rule out this
possibility directly at the wavevectors of interest. First, we 
used an unpolarized beam to monitor the phonon
intensity in this energy range at high momentum transfers where
lattice dynamical models \cite{fong} predict strong phonon scattering.
No difference of the intensities above and below $\rm T_c$ was observed
outside the statistical error. 

Second, we used spin polarization analysis
to separate the non-spin-flip (phonon) and spin-flip (magnetic) 
cross sections. Depending on the energy transfer, the data were taken
in two scattering geometries in which momentum 
transfers of the forms either (H,H,L) or (3H,H,L) were accessible. 
Use of both configurations was necessary to
avoid accidental elastic scattering which
can give rise to spurious signals at certain energy transfers \cite{fong}. 
Typical raw data are shown in Fig. 1b. While at energies around 30-45 meV
the phonon cross section is peaked around ${\bf Q}=(\frac{1}{2},\frac{1}{2})$
\cite{fong}, it does not depend strongly on the in-plane momentum transfer
at lower energies, as demonstrated
for $\hbar \omega = 21$ meV in the figure. The spin-flip scattering is
peaked at ${\bf Q}=(\frac{1}{2},\frac{1}{2})$ and 
shows the characteristic factor-of-two
polarization dependence when the neutron spin at the sample position
is rotated from {\bf Q}-parallel (horizontal field)
to {\bf Q}-perpendicular (vertical field) \cite{moon}. 
The peak in the unpolarized-beam data of Fig. 1a 
thus arises exclusively from magnetic scattering.
The filled circles in Fig. 2 are the spin-flip peak intensity at 25 meV, 
corrected for the background and measured with counting statistics
comparable to those of the unpolarized-beam data. The non-spin-flip scattering
was measured with similar statistics and showed no change upon cooling
through $\rm T_c$. The
excellent agreement with the unpolarized-beam data again confirms the magnetic
origin of the enhanced cross section below $\rm T_c$.

Fig. 3 shows the difference between spectra measured above and below
$\rm T_c$ for both $\rm YBa_2 Cu_3 O_{6.5}$ and $\rm YBa_2 Cu_3 O_{6.7}$.
In addition to these scans, 
{\bf Q}-scans were performed at several energies above and below $\rm T_c$
in order to obtain the energy and temperature dependence of the 
background throughout the Brillouin zone. For most of the data in the figure
the background (originating mostly from single-phonon and multiphonon scattering
events) was found to be temperature independent in the temperature range of
interest. For the measurements taken below 18 meV in $\rm YBa_2 Cu_3 O_{6.7}$
the background increases uniformly throughout the Brillouin zone by about
10\% upon heating from 10K to 80K. The origin of this increase is unknown
(though it could arise from multiphonon scattering), but it is unlikely to be
related to magnetic excitations. However, the magnitude of the increase is
too small to definitively rule out a magnetic origin by polarization analysis.
The subtractions below 18 meV in the lower panel of Fig. 3 were 
corrected for this overall effect and thus show only differences centered at
${\bf Q} = (\frac{1}{2},\frac{1}{2})$. 
The data were further corrected for the Bose population factor 
$[1-\exp(-\hbar \omega / k_B T)]^{-1}$ in order to convert the
magnetic cross section to the dynamical spin susceptibility 
$\chi''({\bf q},\omega)$. Following the same procedure as
discussed above for $\rm YBa_2 Cu_3 O_{6.5}$, we found that the 33 meV spectral
weight enhancement for $\rm YBa_2 Cu_3 O_{6.7}$ again sets in below
$\rm T_c$. The difference plots of Fig. 3 therefore directly reveal the
the effect of superconductivity on $\chi''({\bf q},\omega)$.

Several interesting features are apparent in the difference spectra.
First, the $\rm YBa_2 Cu_3 O_{6.5}$ data show that most (but not
all) of the spectral weight enhancement around 25 meV is drawn from 
higher energies in the 30-40 meV range. In the present study 
we used a relatively coarse energy resolution in order to maximize
the signal at high energies and were therefore confined to energies above
7 meV. Previous work with tighter resolution \cite{grenoble}
shows that the spectral weight is suppressed at lower energies in
the superconducting state, thus accounting for the remaining
spectral weight. The conservation of spectral weight is expected on 
the basis of the total moment sum rule \cite{lovesey}, but the
fact that the resonance spectral weight is drawn from
normal-state excitations of both higher and lower energy is
surprising and inconsistent with the simplest picture in which
the resonance is exclusively built up from states below the
superconducting energy gap.

The situation is somewhat different for $\rm YBa_2 Cu_3 O_{6.7}$. Below
12 meV intensity centered around ${\bf Q} = (\frac{1}{2},\frac{1}{2})$
is observed {\it neither} in the superconducting state 
{\it nor} in the normal state immediately
above $\rm T_c$ (T=80K). Because of the nonzero instrumental resolution this
translates into a normal state gap of $\sim$17 meV, in agreement with
previous work at this doping level \cite{grenoble}. After adjusting for
the background and the thermal population factor, the magnetic intensity
in the relatively narrow energy interval between the normal state gap and
the tail of the 33 meV resonance peak is not strongly
affected by superconductivity
outside of the experimental error. Though the evidence is weaker than for
$\rm YBa_2 Cu_3 O_{6.5}$, the intensity at energies above the resonance
peak again appears to be suppressed in the superconducting state.
Detailed theoretical work is necessary in order to
relate these observations to those made with a variety of
other techniques in the underdoped cuprates \cite{ong}.

The data of Fig. 3 also demonstrate that the 40 meV resonance peak
observed in the optimally doped compound evolves continuously with
the carrier concentration. A synopsis of the data at all three doping levels
is given in Fig. 4. The resonance is broadened in the deeply underdoped regime
but already resolution limited for $\rm YBa_2 Cu_3 O_{6.7}$. Although
the data are of course too sparse to establish a functional dependence
of the resonance energy on $\rm T_c$ or doping level, the resonance
energy increases monotonically with $\rm T_c$, and the presently available 
data are consistent with a simple proportionality. A very recent
photoemission study of underdoped $\rm Bi_2 Sr_2 Ca Cu_2 O_{8+\delta}$
\cite{harris} demonstrates that the superconducting energy gap in
the single-particle density of states is independent of doping. A
reconciliation of this observation with our study of the neutron
peak presents a challenge to models in which both phenomena are
directly related [4-12].

The data of Fig. 4 also dispel possible speculations about an essential
relationship between the 40 meV resonance and a phonon of slightly
higher energy (42.5 meV) and similar dynamical structure factor observed in
$\rm YBa_2 Cu_3 O_7$ \cite{fong}. This proximity must now be regarded
as purely coincidental. Since the 42.5 meV phonon does not involve the
chain oxygens, it can be used to calibrate the absolute spectral weight
of the resonance at different doping levels. A preliminary normalization
indicates that for both samples the resonance spectral weights are 
within a factor of two of $\int \, d(\hbar \omega) \, \chi_{\rm res}'' 
({\bf q},\omega) = 0.5$ found
in $\rm YBa_2 Cu_3 O_7$ \cite{fong}. Details will be given in a forthcoming
full publication \cite{fong1}. Note that the normal state magnetic
intensity in the range 10 meV $\leq$ $\hbar \omega$ $\leq$ 40 meV decreases
dramatically (at least by a factor of five) in the same doping interval.

In summary, our data are consistent with previous work
on underdoped $\rm YBa_2 Cu_3 O_{6+x}$ in the normal 
state \cite{grenoble,tranquada2}, but additional
spectral enhancements, undoubtedly analogs of the
40 meV resonance in $\rm YBa_2 Cu_3 O_7$, are observed in
the superconducting state. Some new and surprising aspects of the
resonance are elucidated, including in particular its doping dependence 
and its relation to the magnetic spectrum in the normal state. 
Obviously, the theoretical models proposed to explain
the 40 meV mode should now be tested against this more extensive data set.

\vspace{.1in}

\noindent {\bf Acknowledgments}

\noindent We are grateful to the Brookhaven neutron scattering group for their
gracious hospitality, and to P. Bourges, S. Chakravarty,
V.J. Emery, A.J. Millis, H. Monien, G. Shirane,
J.M. Tranquada, S.C. Zhang, and especially P.W. Anderson for
helpful discussions and suggestions throughout
this project. The work at Princeton University was supported 
by the MRSEC program of the National
Science Foundation under grant No. DMR94-00362, and by the Packard and
Sloan Foundations. The work at Brookhaven was supported by the US DOE
under contract No. DE-AC02-76CH00016.

\clearpage
\section*{Figure Captions}
\begin{enumerate}
\item
(a) Unpolarized-beam scans at $\hbar \omega = 25$ meV and (b) polarized-beam
scans at $\hbar \omega = 21$ meV for $\rm YBa_2 Cu_3 O_{6.5}$, with
$\rm {\bf Q} = (H,H,-5.4)$. In (a) the
lines are fits to Gaussians as discussed in the text. In (b) the open
(closed) symbols are data taken for horizontal (vertical) field 
at the sample position. The upper profile in (b) shows magnetic scattering
in the spin-flip (SF) channel, while the lower profile is phonon
scattering in the non-spin-flip (NSF) channel. The curves are guides-to-the-eye.
\item
The open circles are the amplitudes of the magnetic cross section at 
$\hbar \omega = 25$meV, extracted from Gaussian fits to unpolarized-beam 
constant-energy scans. The closed circles are the spin-flip peak intensities 
measured at ${\bf Q}=(\frac{3}{2},\frac{1}{2},-1.7)$ with a polarized beam,
corrected for the background and scaled to the unpolarized-beam data.
The inset shows the field-cooled DC susceptibility at H=10G 
measured by SQUID magnetometry
on a small piece cut from the inside (not the surface) of the sample.
Because of demagnetization effects the Meissner fraction was nominally
$>100$\%. The data were therefore normalized to the maximum susceptibility.
\item
Difference between the low temperature ($\rm <T_c$) and high temperature
($\rm \gsim T_c$) magnetic susceptibility at
${\bf Q}=(\frac{1}{2},\frac{1}{2},-5.4)$. The susceptibility was
obtained by adjusting the magnetic cross section for the Bose 
factor and for a weakly temperature-dependent featureless background 
at low energies in the lower panel.
\item
Energy of maximum superconductivity-induced spectral enhancement as a function
of the transition temperature $\rm T_c$. For $\rm YBa_2 Cu_3 O_{6.7}$ and
$\rm YBa_2 Cu_3 O_7$ the enhancement is resolution-limited in energy, and the
error bars are upper bounds on the intrinsic width of the resonance.
\end{enumerate}
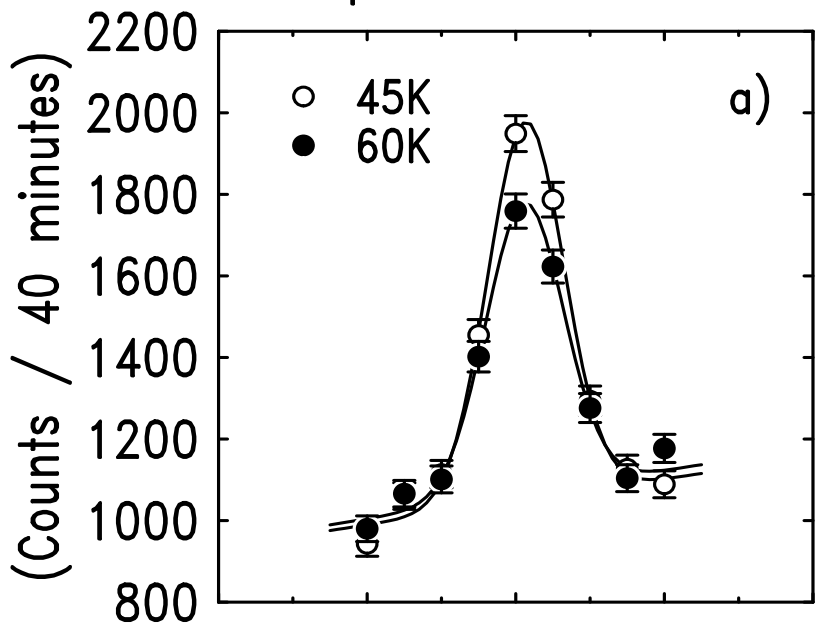
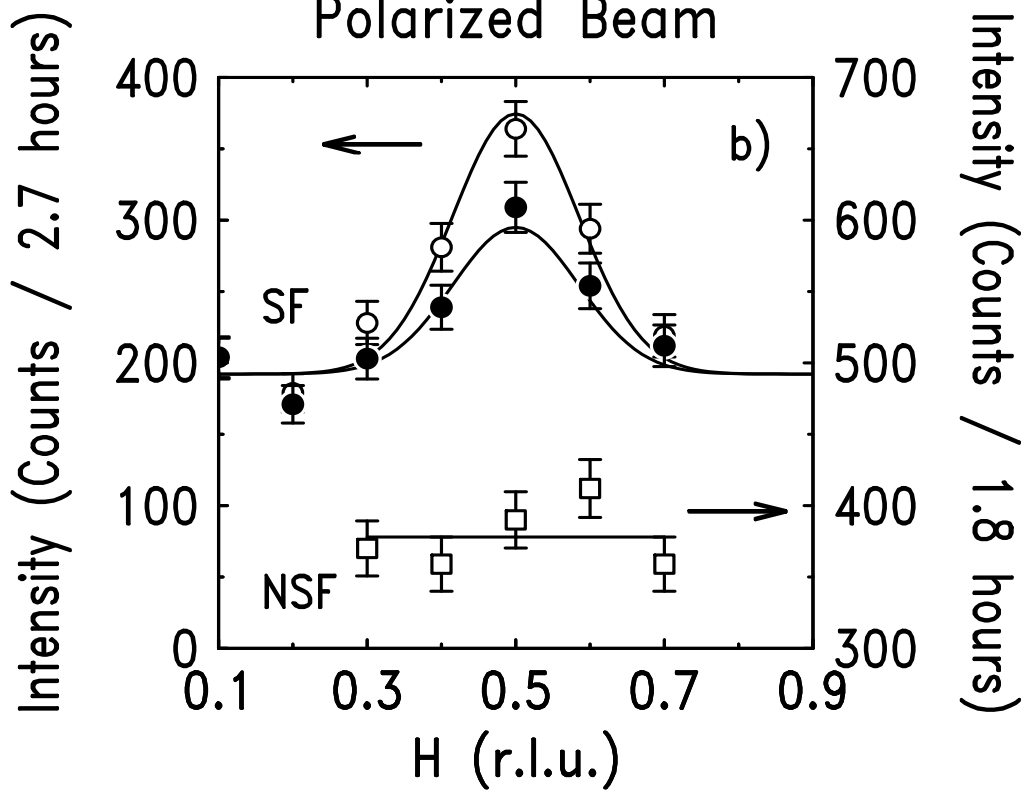
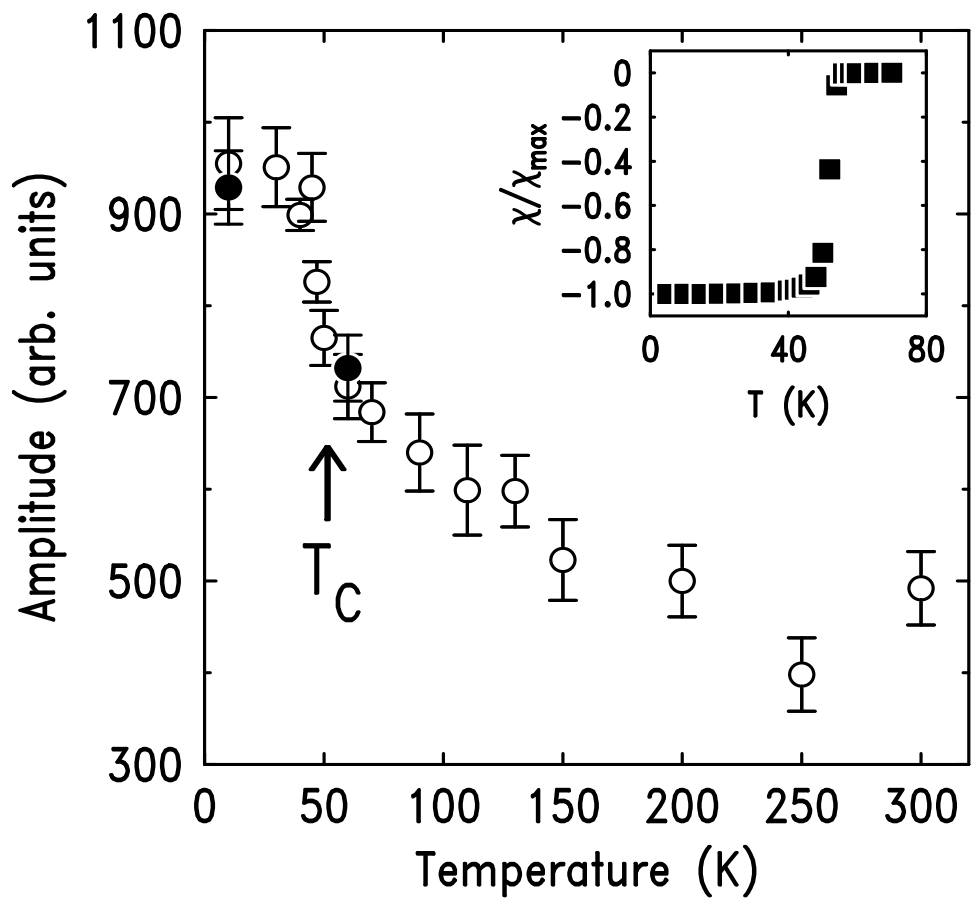
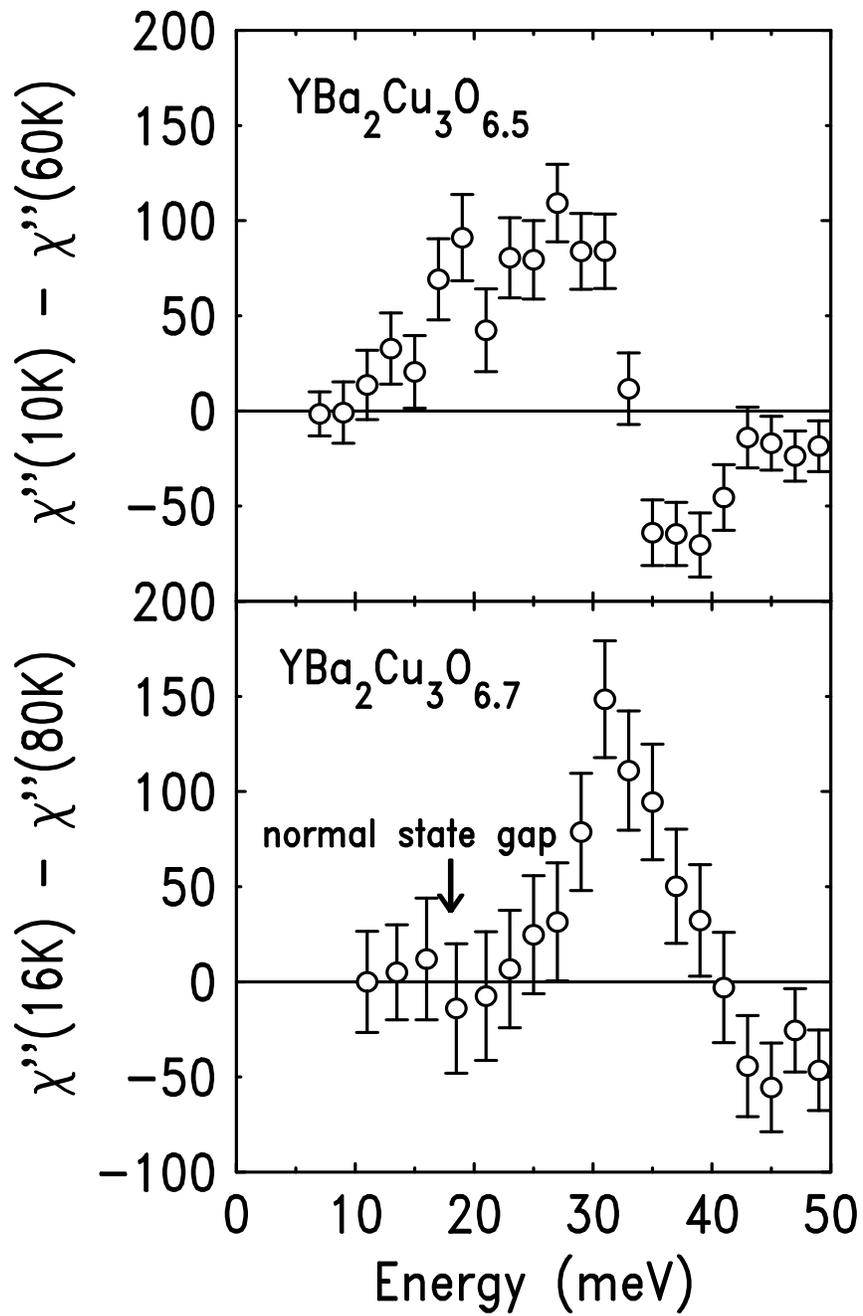
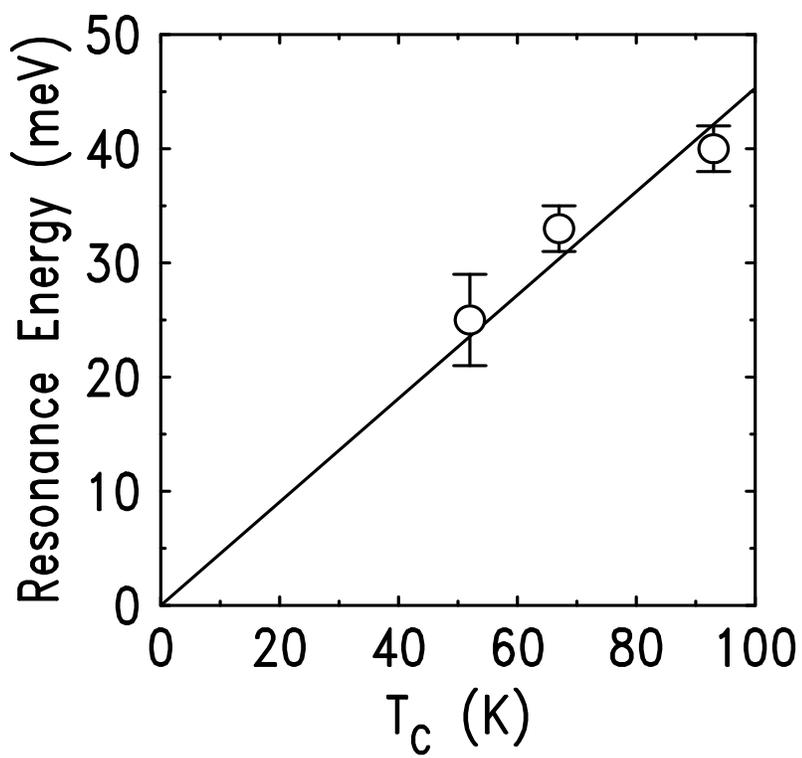
\end{document}